
\input harvmac.tex



\def\unlockat{\catcode`\@=11}
\def\lockat{\catcode`\@=12}

\unlockat

\def\newsec#1{\global\advance\secno by1\message{(\the\secno. #1)}
\global\subsecno=0\global\subsubsecno=0\eqnres@t\noindent
{\bf\the\secno. #1}
\writetoca{{\secsym} {#1}}\par\nobreak\medskip\nobreak}
\global\newcount\subsecno \global\subsecno=0
\def\subsec#1{\global\advance\subsecno
by1\message{(\secsym\the\subsecno. #1)}
\ifnum\lastpenalty>9000\else\bigbreak\fi\global\subsubsecno=0
\noindent{\it\secsym\the\subsecno. #1}
\writetoca{\string\quad {\secsym\the\subsecno.} {#1}}
\par\nobreak\medskip\nobreak}
\global\newcount\subsubsecno \global\subsubsecno=0
\def\subsubsec#1{\global\advance\subsubsecno by1
\message{(\secsym\the\subsecno.\the\subsubsecno. #1)}
\ifnum\lastpenalty>9000\else\bigbreak\fi
\noindent\quad{\secsym\the\subsecno.\the\subsubsecno.}{#1}
\writetoca{\string\qquad{\secsym\the\subsecno.\the\subsubsecno.}{#1}}
\par\nobreak\medskip\nobreak}

\def\subsubseclab#1{\DefWarn#1\xdef
#1{\noexpand\hyperref{}{subsubsection}%
{\secsym\the\subsecno.\the\subsubsecno}%
{\secsym\the\subsecno.\the\subsubsecno}}%
\writedef{#1\leftbracket#1}\wrlabeL{#1=#1}}
\lockat

\def\IL{\relax{\rm I\kern-.18em L}}
\def\IH{\relax{\rm I\kern-.18em H}}
\def\IR{\relax{\rm I\kern-.18em R}}
\def\IC{\relax\hbox{$\inbar\kern-.3em{\rm C}$}}
\def\IZ{\relax\ifmmode\mathchoice
{\hbox{\cmss Z\kern-.4em Z}}{\hbox{\cmss Z\kern-.4em Z}}
{\lower.9pt\hbox{\cmsss Z\kern-.4em Z}}
{\lower1.2pt\hbox{\cmsss Z\kern-.4em Z}}\else{\cmss Z\kern-.4em
Z}\fi}

\def\CN {{\cal N}}

\def\CJ {{\cal J}}
\def\CP {{\cal P }}

\def\CG {{\cal G}}


\def\CN {{\cal N}}

\def\CP {{\cal P }}

\font\manual=manfnt \def\dbend{\lower3.5pt\hbox{\manual\char127}}

\def\IZ{\relax\ifmmode\mathchoice
{\hbox{\cmss Z\kern-.4em Z}}{\hbox{\cmss Z\kern-.4em Z}}
{\lower.9pt\hbox{\cmsss Z\kern-.4em Z}}
{\lower1.2pt\hbox{\cmsss Z\kern-.4em Z}}\else{\cmss Z\kern-.4em
Z}\fi}

\def\p{\partial}
\def\pb{\bar{\partial}}

\def\CJ {{\cal J}}

\def\CN {{\cal N}}

\def\CP {{\cal P }}


\def\Aut{{\rm Aut}}

\def\IZ{\relax\ifmmode\mathchoice
{\hbox{\cmss Z\kern-.4em Z}}{\hbox{\cmss Z\kern-.4em Z}}
{\lower.9pt\hbox{\cmsss Z\kern-.4em Z}}
{\lower1.2pt\hbox{\cmsss Z\kern-.4em Z}}\else{\cmss Z\kern-.4em
Z}\fi}
\def\IB{\relax{\rm I\kern-.18em B}}
\def\IC{{\relax\hbox{$\inbar\kern-.3em{\rm C}$}}}
\def\ID{\relax{\rm I\kern-.18em D}}
\def\IE{\relax{\rm I\kern-.18em E}}
\def\IF{\relax{\rm I\kern-.18em F}}
\def\IG{\relax\hbox{$\inbar\kern-.3em{\rm G}$}}
\def\IGa{\relax\hbox{${\rm I}\kern-.18em\Gamma$}}
\def\IH{\relax{\rm I\kern-.18em H}}
\def\II{\relax{\rm I\kern-.18em I}}
\def\IK{\relax{\rm I\kern-.18em K}}
\def\IP{\relax{\rm I\kern-.18em P}}

\def\lieg{{\underline{\bf g}}}

\def\inbar{\,\vrule height1.5ex width.4pt depth0pt}
\def\Map{{\rm Map}}

\def\p{\partial}
\def\pb{{\bar \p}}

\def\Pic{{\rm Pic}}

\font\cmss=cmss10 \font\cmsss=cmss10 at 7pt
\def\IR{\relax{\rm I\kern-.18em R}}

\def\Tr{\rm Tr}


\def\boxit#1{\vbox{\hrule\hbox{\vrule\kern8pt
\vbox{\hbox{\kern8pt}\hbox{\vbox{#1}}\hbox{\kern8pt}}
\kern8pt\vrule}\hrule}}
\def\mathboxit#1{\vbox{\hrule\hbox{\vrule\kern8pt\vbox{\kern8pt
\hbox{$\displaystyle #1$}\kern8pt}\kern8pt\vrule}\hrule}}


\def\lieg{{\underline{\bf g}}}

\def\inbar{\,\vrule height1.5ex width.4pt depth0pt}

\def\p{\partial}

\def\pb{{\bar \p}}

\font\cmss=cmss10 \font\cmsss=cmss10 at 7pt
\def\IR{\relax{\rm I\kern-.18em R}}

\def\Tr{\rm Tr}


\def\a1{{\cal A}^{1,1}}

%
\lref\blzh{A. Belavin, V. Zakharov, ``Yang-Mills

Equations as inverse scattering problem''

Phys. Lett. B73, (1978) 53}
\lref\afs{Alekseev, Faddeev, Shatashvili,  }
\lref\mickelsson{ J. Mickelsson, ``Kac-Moody groups, topology of

the Dirac  determinant bundle and fermionization,''
Commun. Math. Phys. {\bf 110}(1987)173.}
\lref\bost{L. Alvarez-Gaume, J.B. Bost , G. Moore, P. Nelson, C.
Vafa,
``Bosonization on higher genus Riemann surfaces,''
Commun.Math.Phys.112:503,1987}
\lref\agmv{L. Alvarez-Gaum\'e,
C. Gomez, G. Moore,
and C. Vafa, ``Strings in the Operator Formalism,''
Nucl. Phys. {\bf 303}(1988)455}
\lref\atiyah{M. Atiyah, ``Green's Functions for
Self-Dual Four-Manifolds,'' Adv. Math. Suppl.
{\bf 7A} (1981)129}

\lref\AHS{M.~ Atiyah, N.~ Hitchin and I.~ Singer, ``Self-Duality in
Four-Dimensional
Riemannian Geometry", Proc. Royal Soc. (London) {\bf A362} (1978)
425-461.}
\lref\fmlies{M. F. Atiyah and I. M. Singer,
``The index of elliptic operators IV,'' Ann. Math. {\bf 93}(1968)119}
\lref\bagger{E. Witten and J. Bagger, Phys. Lett.
{\bf 115B}(1982)202}
\lref\banks{T. Banks, ``Vertex Operators in 2D Dimensions,''
hep-th/9503145   }
\lref\berk{N. Berkovits, ``Super-Poincare Invariant Superstring Field
Theory''
hep-th/9503099 }
\lref\biquard{O. Biquard, ``Sur les fibr\'es paraboliques
sur une surface complexe,'' to appear in J. Lond. Math.
Soc.}
\lref\bjsv{hep-th/9501096,
Topological Reduction of 4D SYM to 2D $\sigma$--Models,
 M. Bershadsky, A. Johansen, V. Sadov and C. Vafa }
\lref\BlThlgt{M.~ Blau and G.~ Thompson, ``Lectures on 2d Gauge
Theories: Topological Aspects and Path
Integral Techniques", Presented at the
Summer School in Hogh Energy Physics and
Cosmology, Trieste, Italy, 14 Jun - 30 Jul
1993, hep-th/9310144.}
\lref\bpz{A.A. Belavin, A.M. Polyakov, A.B. Zamolodchikov,
``Infinite conformal symmetry in two-dimensional quantum
field theory,'' Nucl.Phys.B241:333,1984}
\lref\braam{P.J. Braam, A. Maciocia, and A. Todorov,
``Instanton moduli as a novel map from tori to
K3-surfaces,'' Inven. Math. {\bf 108} (1992) 419}
\lref\cllnhrvy{Callan and Harvey, Nucl
Phys. {\bf B250}(1985)427}
\lref\CMR{ For a review, see
S. Cordes, G. Moore, and S. Ramgoolam,
`` Lectures on 2D Yang Mills theory, Equivariant
Cohomology, and Topological String Theory,''
Lectures presented at the 1994 Les Houches Summer School
 ``Fluctuating Geometries in Statistical Mechanics and Field
Theory.''
and at the Trieste 1994 Spring school on superstrings.
hep-th/9411210, or see http://xxx.lanl.gov/lh94}
\lref\devchand{Ch. Devchand and V. Ogievetsky,
``Four dimensional integrable theories,'' hep-th/9410147}
\lref\devchandi{
Ch. Devchand and A.N. Leznov,
``B \"acklund transformation for supersymmetric self-dual theories
for
semisimple
gauge groups and a hierarchy of $A_1$ solutions,'' hep-th/9301098,
Commun. Math. Phys. {\bf 160} (1994) 551}
\lref\dnld{S. Donaldson, ``Anti self-dual Yang-Mills
connections over complex  algebraic surfaces and stable
vector bundles,'' Proc. Lond. Math. Soc,
{\bf 50} (1985)1}

\lref\DoKro{S.K.~ Donaldson and P.B.~ Kronheimer,
{\it The Geometry of Four-Manifolds},
Clarendon Press, Oxford, 1990.}
\lref\donii{
S. Donaldson, Duke Math. J. , {\bf 54} (1987) 231. }

\lref\elitzur{S. Elitzur, G. Moore,
A. Schwimmer, and N. Seiberg,
``Remarks on the Canonical Quantization of the Chern-Simons-
Witten Theory,'' Nucl. Phys. {\bf B326}(1989)108;
G. Moore and N. Seiberg,
``Lectures on Rational Conformal Field Theory,''
, in {\it Strings'89},Proceedings
of the Trieste Spring School on Superstrings,
3-14 April 1989, M. Green, et. al. Eds. World
Scientific, 1990}
\lref\etingof{P.I. Etingof and I.B. Frenkel,
``Central Extensions of Current Groups in
Two Dimensions,'' Commun. Math.
Phys. {\bf 165}(1994) 429}

\lref\evans{M. Evans, F. G\"ursey, V. Ogievetsky,
``From 2D conformal to 4D self-dual theories:
Quaternionic analyticity,''
Phys. Rev. {\bf D47}(1993)3496}
\lref\fs{L. Faddeev and S. Shatashvili, Theor. Math. Fiz., 60 (1984)
206, L. Faddeev, Phys. Lett. B145 (1984) 81, J. Mickelsson, CMP, 97
(1985) 361,
A. Reiman, M. Semenov-Tian-Shansky, L. Faddeev, Funct. Anal. and
Appl.,vol.18,No.4(1984)319}
\lref\fz{I. Frenkel, I. Singer, unpublished.}

\lref\fk{I. Frenkel and B. Khesin, ``Four dimensional
realization of two dimensional current groups,'' Yale
preprint, July 1995.}
\lref\galperin{A. Galperin, E. Ivanov, V. Ogievetsky,
E. Sokatchev, Ann. Phys. {\bf 185}(1988) 1}
\lref\gwdzki{K. Gawedzki, ``Topological Actions in Two-Dimensional
Quantum Field Theories,'' in {\it Nonperturbative
Quantum Field Theory}, G. 't Hooft, A. Jaffe, et. al. , eds. ,
Plenum 1988}
\lref\gmps{A. Gerasimov, A. Morozov, M. Olshanetskii,
 A. Marshakov, S. Shatashvili ,``
Wess-Zumino-Witten model as a theory of
free fields,'' Int. J. Mod. Phys. A5 (1990) 2495-2589
 }
\lref\gerasimov{A. Gerasimov, ``Localization in
GWZW and Verlinde formula,'' hepth/9305090}
\lref\ginzburg{V. Ginzburg, M. Kapranov, and E. Vasserot,
``Langlands Dualtiy for Surfaces,'' IAS preprint}
\lref\giveon{hep-th/9502057,
 S-Duality in N=4 Yang-Mills Theories with General Gauge Groups,
 Luciano Girardello, Amit Giveon, Massimo Porrati, and Alberto
Zaffaroni
}

\lref\gottsh{L. Gottsche, Math. Ann. 286 (1990)193}
\lref\gothuy{L. G\"ottsche and D. Huybrechts,
``Hodge numbers of moduli spaces of stable
bundles on $K3$ surfaces,'' alg-geom/9408001}
\lref\GrHa{P.~ Griffiths and J.~ Harris, {\it Principles of
Algebraic
geometry},
p. 445, J.Wiley and Sons, 1978. }
\lref\ripoff{I. Grojnowski, ``Instantons and
affine algebras I: the Hilbert scheme and
vertex operators,'' alg-geom/9506020.}
\lref\adhmfk{I. Grojnowski,
A. Losev, G. Moore, N. Nekrasov, S. Shatashvili,
``ADHM and the Frenkel-Kac construction,'' in preparation}

\lref\hitchin{N. Hitchin, ``Polygons and gravitons,''
Math. Proc. Camb. Phil. Soc, (1979){\bf 85} 465}

\lref\hklr{Hitchin, Karlhede, Lindstrom, and Rocek,
``Hyperkahler metrics and supersymmetry,''
Commun. Math. Phys. {\bf 108}(1987)535}
\lref\hirz{F. Hirzebruch and T. Hofer, Math. Ann. 286 (1990)255}
\lref\hms{hep-th/9501022,
 Reducing $S$- duality to $T$- duality, J. A. Harvey, G. Moore and A.
Strominger}
\lref\johansen{A. Johansen, ``Infinite Conformal
Algebras in Supersymmetric Theories on
Four Manifolds,'' hep-th/9407109}
\lref\kronheimer{P. Kronheimer, ``The construction of ALE spaces as
hyper-kahler quotients,'' J. Diff. Geom. {\bf 28}1989)665}
\lref\kricm{P. Kronheimer, ``Embedded surfaces in
4-manifolds,'' Proc. Int. Cong. of
Math. (Kyoto 1990) ed. I. Satake, Tokyo, 1991}

\lref\KN{Kronheimer and Nakajima,  ``Yang-Mills instantons
on ALE gravitational instantons,''  Math. Ann.
{\bf 288}(1990)263}
\lref\krmw{P. Kronheimer and T. Mrowka,
``Gauge theories for embedded surfaces I,''
Topology {\bf 32} (1993) 773,
``Gauge theories for embedded surfaces II,''
preprint.}

\lref\hypvol{A. Losev, G. Moore, N. Nekrasov, S. Shatashvili,
``Localization for Hyperkahler Quotients,
Integration over Instanton Moduli,
and ALE Spaces,'' in preparation}
\lref\fdrcft{A. Losev, G. Moore, N. Nekrasov, S. Shatashvili, in
preparation.}
\lref\lmns{A. Losev, G. Moore, N. Nekrasov, and
S. Shatashvili,  ``Four-Dimensional Avatars of
Two-Dimensional RCFT,''  hep-th/9509151.}

\lref\maciocia{A. Maciocia, ``Metrics on the moduli
spaces of instantons over Euclidean 4-Space,''
Commun. Math. Phys. {\bf 135}(1991) , 467}
\lref\mick{I. Mickellson, CMP, 97 (1985) 361.}

\lref\milnor{J. Milnor, ``A unique decomposition
theorem for 3-manifolds,'' Amer. Jour. Math, (1961) 1}
\lref\taming{G. Moore and N. Seiberg,
``Taming the conformal zoo,'' Phys. Lett.
{\bf 220 B} (1989) 422}
\lref\nair{V.P.Nair, ``K\"ahler-Chern-Simons Theory'', hep-th/9110042
}
\lref\ns{V.P. Nair and Jeremy Schiff,
``Kahler Chern Simons theory and symmetries of
antiselfdual equations'' Nucl.Phys.B371:329-352,1992;
``A Kahler Chern-Simons theory and quantization of the
moduli of antiselfdual instantons,''
Phys.Lett.B246:423-429,1990,
``Topological gauge theory and twistors,''
Phys.Lett.B233:343,1989}
\lref\nakajima{H. Nakajima, ``Homology of moduli
spaces of instantons on ALE Spaces. I'' J. Diff. Geom.
{\bf 40}(1990) 105; ``Instantons on ALE spaces,
quiver varieties, and Kac-Moody algebras,'' preprint,
``Gauge theory on resolutions of simple singularities
and affine Lie algebras,'' preprint.}
\lref\nakheis{H.Nakajima, ``Heisenberg algebra and Hilbert schemes of
points on
projective surfaces ,'' alg-geom/9507012}
\lref\ogvf{H. Ooguri and C. Vafa, ``Self-Duality
and $N=2$ String Magic,'' Mod.Phys.Lett. {\bf A5} (1990) 1389-1398;
``Geometry
of$N=2$ Strings,'' Nucl.Phys. {\bf B361}  (1991) 469-518.}
\lref\park{J.-S. Park, ``Holomorphic Yang-Mills theory on compact
Kahler
manifolds,'' hep-th/9305095; Nucl. Phys. {\bf B423} (1994) 559;
J.-S.~ Park, ``$N=2$ Topological Yang-Mills Theory on Compact
K\"ahler
Surfaces", Commun. Math, Phys. {\bf 163} (1994) 113;
J.-S.~ Park, ``$N=2$ Topological Yang-Mills Theories and Donaldson
Polynomials", hep-th/9404009}
\lref\parki{S. Hyun and J.-S. Park,
``Holomorphic Yang-Mills Theory and Variation
of the Donaldson Invariants,'' hep-th/9503036}
\lref\pohl{Pohlmeyer, Commun.
Math. Phys. {\bf 72}(1980)37}
\lref\pwf{A.M. Polyakov and P.B. Wiegmann,
Phys. Lett. {\bf B131}(1983)121}
\lref\prseg{A.~Pressley and G.~Segal, "Loop Groups", Oxford Clarendon
Press, 1986}
\lref\rade{J. Rade, ``Singular Yang-Mills fields. Local
theory I. '' J. reine ang. Math. , {\bf 452}(1994)111; {\it ibid}
{\bf 456}(1994)197; ``Singular Yang-Mills
fields-global theory,'' Intl. J. of Math. {\bf 5}(1994)491.}
\lref\segal{G. Segal, The definition of CFT}
\lref\seiberg{hep-th/9407087,
Monopole Condensation, And Confinement In $N=2$ Supersymmetric
Yang-Mills
Theory, N. Seiberg and E. Witten;
hep-th/9408013,  Nathan Seiberg;
hep-th/9408099,
Monopoles, Duality and Chiral Symmetry Breaking in
N=2 Supersymmetric QCD, N. Seiberg and E. Witten;
hep-th/9408155,
Phases of N=1 supersymmetric gauge theories in four dimensions, K.
Intriligator
and N. Seiberg; hep-ph/9410203,
Proposal for a Simple Model of Dynamical SUSY Breaking, by K.
Intriligator, N.
Seiberg, and S. H. Shenker;
hep-th/9411149,
 Electric-Magnetic Duality in Supersymmetric Non-Abelian Gauge
Theories,
 N. Seiberg; hep-th/9503179 Duality, Monopoles, Dyons, Confinement
and Oblique
Confinement in Supersymmetric $SO(N_c)$ Gauge Theories,
K. Intriligator and N. Seiberg}
\lref\sen{A. Sen,
hep-th/9402032, Dyon-Monopole bound states, selfdual harmonic
forms on the multimonopole moduli space and $SL(2,Z)$
invariance,'' }
\lref\shatashi{S. Shatashvili,
Theor. and Math. Physics, 71, 1987, p. 366}
\lref\thooft{G. 't Hooft , ``A property of electric and
magnetic flux in nonabelian gauge theories,''
Nucl.Phys.B153:141,1979}
\lref\vafa{C. Vafa, ``Conformal theories and punctured
surfaces,'' Phys.Lett.199B:195,1987 }
\lref\VaWi{C.~ Vafa and E.~ Witten, ``A Strong Coupling Test of
$S$-Duality",
hep-th/9408074.}
\lref\vrlsq{E. Verlinde and H. Verlinde,
``Conformal Field Theory and Geometric Quantization,''
in {\it Strings'89},Proceedings
of the Trieste Spring School on Superstrings,
3-14 April 1989, M. Green, et. al. Eds. World
Scientific, 1990}
\lref\mwxllvrld{E. Verlinde, ``Global Aspects of
Electric-Magnetic Duality,'' hep-th/9506011}
\lref\wrdhd{R. Ward, Nucl. Phys. {\bf B236}(1984)381}
\lref\ward{Ward and Wells, {\it Twistor Geometry and
Field Theory}, CUP }
\lref\wittenwzw{E. Witten, ``Nonabelian bosonization in
two dimensions,'' Commun. Math. Phys. {\bf 92} (1984)455 }
\lref\grssmm{E. Witten, ``Quantum field theory,
grassmannians and algebraic curves,'' Commun.Math.Phys.113:529,1988}
\lref\wittjones{E. Witten, ``Quantum field theory and the Jones
polynomial,'' Commun.  Math. Phys.}
\lref\wittentft{E.~ Witten, ``Topological Quantum Field Theory",
Commun. Math. Phys. {\bf 117} (1988) 353.}
\lref\Witdgt{ E.~ Witten, ``On Quantum gauge theories in two
dimensions,''
Commun. Math. Phys. {\bf  141}  (1991) 153.}
\lref\Witfeb{E.~ Witten, ``Supersymmetric Yang-Mills Theory On A
Four-Manifold,'' J. Math. Phys. {\bf 35} (1994) 5101.}
\lref\Witr{E.~ Witten, ``Introduction to Cohomological Field
Theories",
Lectures at Workshop on Topological Methods in Physics, Trieste,
Italy,
Jun 11-25, 1990, Int. J. Mod. Phys. {\bf A6} (1991) 2775.}
\lref\wittabl{E. Witten,  ``On S-Duality in Abelian Gauge Theory,''
hep-th/9505186}
\lref\nov{ S. Novikov,"The Hamiltonian
formalism and many-valued analogue of  Morse theory",
Russian Math.Surveys 37:5(1982),1-56;
E. Witten, ``Global Aspects of Current Algebra,''
Nucl.Phys.B223:422,1983}
\lref\faddeevlmp{ L. D. Faddeev, ``Some Comments on Many Dimensional
Solitons'',
Lett. Math. Phys., 1 (1976) 289-293.}

\Title{ \vbox{\baselineskip12pt\hbox{hep-th/9511185}
\hbox{PUPT-1568}
\hbox{ITEP-TH.10/95}
\hbox{YCTP-P18-95}}}
{\vbox{
\centerline{Central Extensions  }
\centerline{  of }
\centerline{Gauge Groups
Revisited}}}\footnote{}
\medskip
\centerline{Andrei Losev $^1$, Gregory Moore $^2$,
Nikita Nekrasov $^3$, and Samson Shatashvili $^{4}$\footnote{*}{On
leave of
absence from St. Petersburg Steklov Mathematical Institute, St.
Petersburg,
Russia.}}

\vskip 0.5cm
\centerline{$^{1,3}$ Institute of Theoretical and Experimental
Physics,
117259, Moscow, Russia}
\centerline{$^3$ Department of Physics,
Princeton University, Princeton NJ 08544}
\centerline{$^{1,2,4}$ Dept.\ of Physics, Yale University,
New Haven, CT  06520, Box 208120}
\vskip 0.1cm
\centerline{losev@genesis5.physics.yale.edu}
\centerline{moore@castalia.physics.yale.edu}
\centerline{nikita@puhep1.princeton.edu}
\centerline{samson@euler.physics.yale.edu}

\medskip
\noindent
We present
an explicit construction for the central extension of the
group $\Map(X, G)$ where $X$ is  a compact manifold
and $G$ is a Lie group.
If $X$ is a complex  curve  we obtain a simple construction of
the extension by the Picard variety $\Pic(X)$.
The construction is easily adapted to the extension
of $\Aut(E)$, the gauge group of
automorphisms of a nontrivial vector bundle $E$.

\Date{November 26, 1995; Revised December 19,1995}

\newsec{Statement of the problem}

This paper continues an investigation into the
application and generalization of two-dimensional
field theory to higher dimensional theories
begun in  \lmns.  Here we explain how the
higher dimensional analog of the central
charge of a current algebra  is ``integrated''  to
the central  extension of the corresponding
group.

Let $X$ be an $n$-dimensional compact manifold,
$G$ a Lie group and ${\CG} = \Map(X, G)$ the space of
differentiable maps. It is well-known that
when $G$ is simple and simply connected
the covering group of $\CG$ has a universal
central extension by
\eqn\extdsp{
\CJ = \Omega^1(X;\IR)/Z^1_{\IZ}(X)
}
where $\Omega^j(X)$ is the space of all
differentiable $j$-forms on $X$, $Z^j(X)$
is the space of closed $j$-forms, and
$Z^j_{\IZ}$ is the space of closed forms with
integral periods \prseg.
Correspondingly, the Lie algebra is
extended by the space:
\eqn\extalg{
J=
 \Omega^1(X)/d\Omega^0(X)  \cong  {Z}^{n-1}(X)^{\vee}
}
by means of  the cocycle:
\foot{$\Tr$  is normalized so that, if
$\tilde G$ is the simply connected cover, an
integral generator of $H^4(B\tilde G;\IZ)$ is
defined by ${ 1 \over  8 \pi^2} {\Tr} F^2.$ }
\eqn\alcoc{\langle c(X,Y) , \alpha \rangle =
{ 1 \over  8 \pi^2}  \int_{X}
\alpha \wedge {\Tr} (X d Y)}
for $\alpha\in Z^{n-1}(X)$ and $X,Y\in \Omega^0(X;\lieg)$.

It has been
emphasized in \etingof\  that it would be
desirable to make the abstract construction
of the universal central
extension $\widehat{\CG}$  more explicit.
In this note we give such a construction.
It is similar to Mickelsson's approach
 \mickelsson\
for the case $X = S^{1}$  and follows the ideas of
section (4.4) of \prseg.
A different solution to this problem,
for the case when $X$ is  a Riemann surface,
 was recently proposed in \fk.

A slight generalization of the above problem replaces  the
group $\Map(X,G)$ by the group
$\Aut(E)$ of gauge transformations of a principal $G$-bundle
$E$ over $X$. In order to write down the
Lie algebra
 cocycle one fixes a connection $\nabla$ in the adjoint
bundle $ad(E)$ and defines:
\foot{Under  a
change of connection  by an $ad(E)$-valued one form $A$ the
cocycle  changes by a  coboundary:
$c_{\nabla + A}  -c_{\nabla}=\delta \epsilon_{A} ,$
where
$ \epsilon_{A} (X)= { 1 \over  8 \pi^2} {\Tr} (XA) . $}
\eqn\galg{c_\nabla(X,Y) = { 1 \over  8 \pi^2} {\Tr} ( X {\nabla} Y)\qquad . }
Our construction generalizes to give the
universal central extension of $\Aut(E)$.

\newsec{General construction}

\subsec{Extension of the universal covering  of  the group $\Map_0(X,G)$}

We begin by constructing the extension of
 the universal  covering
$U\CG$   of the
component of the identity
$\CG_0=\Map(X,G)_0$  of   $\CG=\Map(X,G)$.
If $B \subset A$ and $D \subset C$ we let
$\Map ((A,B); (C,D))$ denote the space of smooth maps $f$ of $A$ to
$C$, such
that $f(B) \subset D$.
Introducing $I=[0,1]$ we define:
\eqn\denfs{
\eqalign{
\CP\CG & \equiv  \Map((X \times I, X \times \{ 1 \} ) ; (G, 1))\cr
{\Omega}{\CG} & \equiv \Map((X \times I, X \times \{ 0, 1 \} ) ; (G, 1))\cr}
}
and let  $ \Omega_0 \CG\subset   \Omega \CG$
be the component of the identity.
The construction is summarized by the
diagram:
\eqn\diagram{
\matrix{
&  &          &  &1 &
 & 1  &  &  \cr
&  &          &  &\uparrow &
 & \uparrow  &  &  \cr
  &                   &                &                    &
     &                      &   &           &  \cr
1 & \rightarrow & \CJ         & \rightarrow& \widehat{U\CG} &
\rightarrow & U\CG
       & \rightarrow & 1\cr
  &                   &                &                    &
     &                      &   &           &  \cr
   &                   & \uparrow &                  & \uparrow
&
             & \uparrow&                   &  \cr
  &                   &                &                    &
     &                      &   &           &  \cr
1 & \rightarrow & \CN         & \rightarrow& \widehat{\CP\CG} &
\rightarrow & \CP\CG   & \rightarrow & 1\cr
  &                   &                &                    &
     &                      &   &           &  \cr
   &                   &                &                    &
       &  \nwarrow  \psi & \uparrow&                &  \cr
   &                   &                &                    &
     &                      &   &           &  \cr
  &                   &                &                    &
     &                      &  \Omega_0 \CG &           &  \cr
  &                   &                &                    &
     &                      & \uparrow &           &  \cr
  &                   &                &                    &
     &                      & 1 &           &  \cr}
}

In the rightmost column of \diagram\
 we represent the group  $U\CG$ as a quotient.
To obtain the middle line we first construct a topologically trivial
extension of  $ \CP \CG $  by the space of two-forms
$\Omega^2(X \times I)$
using the group law:
\eqn\cccgi{
(g_{1}, e_{1} ) \cdot (g_{2}, e_{2}) =
\bigl( g_{1}g_{2}, e_{1} + e_{2}  + C(g_{1}, g_{2})\bigr)
}
where  $C$ is a cocycle given by:
\eqn\cocycle{ C(g_{1}, g_{2})=
{ 1 \over  8 \pi^2}
{\Tr} (g_{1}^{-1} dg_{1} \wedge
dg_{2}
g_{2}^{-1})\qquad . }
We would like to construct an embedding
$\psi$ of $\Omega_0 \CG$ as a normal subgroup of
$\widehat {\CP \CG} $ using
the 3-form $\omega_3={\Tr} (g^{-1} d g)^3$ on the group $G$.
Accordingly, we choose an
extension  $\tilde h(x,t,\tilde t) $
 of $h\in \Omega_0 \CG$
to $X\times I \times \tilde I $ such that
$\tilde h(x,t,\tilde t=1) =1 $ and write:
\eqn\normemb{
\psi(h) \equiv \biggl( h , { 1 \over  24 {\pi}^{2}}
\int_{\tilde I }  \tilde{h}^{*} \omega_3\biggr) \quad .
}
The second entry of
\normemb\ is
an element of $\Omega^2(X\times I)$ which depends
on $\tilde h$.  Two choices of extension lead to
a difference by
the closed two-form:
${1 \over  24 \pi^2}
\int_{S^1 } \bar{h}^{*} \omega_3$.
This two-form has integral periods since for any
cycle $\gamma$
in $ Z_2(X\times I, X\times\{ 0,1\} )$ we have
the corresponding period:
\eqn\wzact{{ 1 \over  24 {\pi}^{2}}
\int_{\gamma \times S^{1}} \bar{h}^{*} \omega_3 \in   {\IZ}.
}
Thus the
difference for two  choices of
extension is an element of
$ Z^2_{\IZ} (X\times I, X\times\{ 0,1\} )$,
the space of closed two-forms vanishing on
$X\times\{ 0,1\}$ and having integral periods.
Hence we must extend $\CP\CG$
in \diagram\ by the quotient
space
$
\CN \equiv \Omega^2(X\times I) / Z^2_{\IZ} (X\times I, X\times\{
0,1\} )$.
Using the Polyakov-Wiegmann formula
\eqn\polwi{
(g_1 g_2)^{*}\omega_3=g_{1}^{*}\omega_3+
g_{2}^{*} \omega_3 -  d(3{\Tr} (g_{1}^{-1} dg_1 \wedge
dg_2
g_{2}^{-1}))}
one easily checks that
$\psi$ is a group homomorphism
and that the image is a normal subgroup.

The  projection in the middle column of \diagram\  is defined
by restriction of $g$ to the boundary and gives
$\widehat{U\CG}\cong \widehat {\CP\CG}/ \psi(\Omega_0 \CG)$.
Correspondingly, we have a map of the centers
$\CN \rightarrow \CJ$ by integration along $I$.

\subsec{ Central extension for non-simply-connected  $\CG$ }

When $\CG$ is
  not  simply connected   we take a composition
of  the above extension of the universal covering  with the extension of
the group $\CG$ by its fundamental group:
\eqn\piext{
1 \rightarrow \pi_1(\CG) \rightarrow U\CG \rightarrow  \CG \rightarrow  1
}
to get the universal central extension of $\CG$:
\eqn\totalc{
\eqalign{
1 \rightarrow  \widehat{\CJ} &
 \rightarrow  \widehat{\CG} \rightarrow \CG \rightarrow1 \cr
 \widehat{\CJ}\equiv &
 \widehat{\Omega\CG}/\psi(\Omega_0 \CG)\qquad . \cr}
}
Here  $ \widehat{\Omega\CG}$ is the  restriction  of  the extension
$\widehat{\CP\CG}$  of  the group  $\CP\CG$
to its subgroup  $\Omega\CG$.
In order  to show that
$ \widehat{\CJ}$  is in the center
of $\widehat{U\CG} $ one must use \polwi\ and
the result that the fundamental group  of {\it any}
Lie group is abelian. In general $ \widehat{\CJ}$
is {\it itself} an extension
\eqn\nwexteni{
1 \rightarrow \CJ \rightarrow  \widehat{\CJ} \rightarrow \pi_1(\CG)
\rightarrow 1 }
If  $\pi_1(\CG)$ has no  torsion  then
$\widehat{\CJ} \cong
\pi_1(\CG) \oplus \CJ $
since the   projection of  an abelian group to  ${\IZ}^{n}$
splits.

\subsec{ Extension of \Aut(E)}

The above construction generalizes to
the gauge group of
a nontrivial  bundle $E$ by making the following
replacements.
\foot{We continue to
assume that $\Aut(E)$ is connected.}
Let $\pi:  X \times I \rightarrow X$ be a
projection. The group $\CP\CG$
is replaced by the subgroup of
$\Aut(\pi^*(E))$ of automorphisms
which are trivial at $t=1$.
$\Omega \CG$ is replaced by the
subgroup of automorphisms which are
trivial at $t=0,1$.

Generalizing the extension in \cccgi\  requires a
choice of connection on $\Aut(E)$ and is
defined by the cocycle:
$c_{\nabla}(g_1,g_2) = { 1 \over  8 \pi^2}
{\Tr} g_1^{-1} \tilde{\nabla} g_1 \tilde{\nabla} g_2 g_2^{-1}.
$
Here $\nabla$ is  a connection on $\Aut(E)$, induced by
connection $d+A$ on $E$, so
$g^{-1} \nabla g \equiv g^{-1} d g + g^{-1} A g - A$,
where  $\tilde {\nabla}$ is a  pullback connection.
The generalization of  the embedding
$\psi$  is given by
\eqn\nwspi{
\psi_\nabla(h) = \Biggl(  h , {1 \over  24 \pi^2} \int_{\tilde I}
\biggl[
{\Tr } (\tilde h^{-1} \tilde{\nabla} \tilde h)^3
+ 3 {\Tr } \ F_{\tilde{\nabla} }[\tilde h^{-1}  \tilde \nabla \tilde h
+
(\tilde \nabla \tilde h)  \tilde h^{-1}  ] \biggr]
\Biggr)\quad .
}
Here we use the 3-form originally discovered in \faddeevlmp\ in connection
with multi-dimensional solitons (for a recent application see \gerasimov).
The extra terms in the second entry
in  \nwspi\ are required in order
for
$\psi_\nabla$ to define a group homomorphism,
or, equivalently, in order to satisfy the
PW formula:
$\psi_\nabla(g_1  g_2)=\psi_\nabla(g_1) +\psi_\nabla(g_2)
+c_{\nabla}(g_1,g_2)$.Again, two
choices of extension of $\tilde h$ lead to an
ambiguity in \nwspi\  by an element of
$Z^2_{\IZ} (X\times I, X\times\{  0,1\} )$
(the periods are related to characteristic
numbers of  vector bundles constructed from
$E$). So, the group $\Aut(E)$ is
extended by the same space $\widehat{\CJ}$.

\subsec{ Extension when  $\Aut(E)$ is not connected}

Now suppose that $\pi_0(\Aut(E))$ is not
trivial so  $\Aut(E)= \amalg_\alpha \Aut(E)_\alpha$.
The construction above generalizes by
letting
$\CP\CG=\amalg \CP\CG_\alpha$ where, for each
component $\alpha$ we choose a standard
element $g^0_\alpha$ (with $g^0_0 = 1$)
and let
$\CP\CG_\alpha = \{ g\in \Aut(E):
g(x,t=1) = g^0_\alpha(x)
\}$.
The group $\Omega_0 \CG$ remains unchanged as
does the construction.

\subsec{Explicit formulas for the central extension}

In order to make the construction more
explicit we must make two choices.
First, for each $\gamma\in \pi_1(\CG)$, we choose a
representative
$L(\gamma)\in \Omega\CG$. Second, we
choose  a  continuation $\Phi(g)$
of $g$, i.e.  a  map of sets $\CG \rightarrow \CP\CG $ that inverts
the restriction  of  $\CP\CG$  to $X$.
In general neither  $L$ nor $\Phi$ is a homomorphism.
Indeed, if  $G$ is simple then
 $\Phi$  cannot be a homomorphism.
Elements of the  centrally extended  group $ \widehat {\CG}$
are  left-cosets
\eqn\elecrlygee{
(g, \Phi(g), 0)(1,L(\gamma),\lambda) \psi(\Omega_0 \CG)  \subset
\widehat{\CP\CG}
}
where $\lambda \in \CJ$.
Elements of the center  $ \widehat{\CJ}=\widehat {\Omega\CG}/ \psi(\Omega_0
\CG)$
of the extension \totalc\
are also cosets:
$(1, L(\gamma), \lambda) \psi(\Omega_0 \CG)
\subset \widehat{\Omega\CG}$.

We now give explicit formulae for the multiplication of
the cosets. $ \widehat{\CJ}$ is itself a central extension
\nwexteni. Thus, as a set it is
the space of pairs  $(\gamma, \lambda)$,
but the  multiplication involves a cocycle:
$(\gamma_1,\lambda_1)(\gamma_2,\lambda_2)=
\bigl( \gamma_1 + \gamma_2, \lambda_1 + \lambda_2 +
CL(\gamma_1,\gamma_2)\bigr)$.
The cocycle $CL$ is defined by choosing a
homotopy $h$ between  the loops \foot{ $*$ stands for the path multiplication.}
 $L(\gamma_1 * \gamma_2)$
and  $L(\gamma_1) * L(\gamma_2)$
and writing:\eqn\centcocy{
CL(\gamma_1,\gamma_2)=
{ 1 \over  24 {\pi}^{2}}\int_{I \times \tilde{I}}h^{*} \omega_3 \qquad. }
Multiplication of  two cosets  \elecrlygee\ in  $\widehat{\CP\CG}$
 leads to the central element
 $(\gamma(g_1, g_2), \lambda(g_1,g_2))$  given by:
\eqn\answer{
\eqalign{
 \gamma(g_1, g_2) =&  [  \varphi_{1,2} ] \in\pi_1(\CG) \cr
 \lambda(g_1,g_2)      =   &\int_{I} C(\Phi(g_1), \Phi(g_2))+
{ 1 \over  24 {\pi}^{2}}\int_{I  \times \tilde{I}}
  \tilde{h}^{*}\omega_3 ,\cr
}}
The loop $\varphi_{1,2} \in  \Omega\CG$  is obtained  by
``glueing together'' the paths  $\Phi(g_1) \cdot  \Phi(g_2)$  and
$\Phi(g_1 \cdot  g_2)$. More precisely, denoting by
$()^{\rm inv}$ the operation of taking the inverse in the semigroup of paths,
$\varphi_{1,2}=(\Phi(g_1) \cdot \Phi(g_2)) * (\Phi(g_1 \cdot g_2))^{\rm inv}$.
  $\tilde{h}$  is a homotopy  in  $\CG$ between
the  loop  $\varphi_{1,2} $  and  the representative
of its  homotopy class  $L([\varphi_{1,2} ])$.

\subsec{Relation to the descent procedure}

In light of the above results it is instructive to reconsider the
descent procedure for constructing  gauge group cocycles
with  values in functionals of gauge connections \fs.
Recall that one  introduces three operations $d,d^{-1}, \delta,$ on
differential-form valued functionals
of group elements and gauge connections such that
$\delta^2=d^2=0$. $\delta$  is a group cochain
differential \fs;
$d^{-1}$ is
defined in \nov\ and is essentially the operation
$\int_I$ used in equation \normemb\  above.
Starting from  a  $2m$-form  $\Omega_{2m} = {\Tr}  F^m$,
which satisfies $\delta\Omega_{2m}=d\Omega_{2m}=0$,
one applies the operation $d^{-1}\delta$ a total of
$k$ times  to get
$\Xi = {\int}_Xd^{-1}\delta\cdots d^{-1}\delta
d^{-1}\Omega_{2m}$,  which satisfies
$\delta \Xi=0$ and hence is a
cocycle of degree $k$ in dimension $2m-k-1$.

Comparing with \alcoc\ it is apparent that
one might have started with an $n+3$-form
$\Omega_{n+3}'={\alpha} \wedge {\Tr}F^2$
 where $\alpha$ is a closed $n-1$-form with
integer periods.
Repeating  the descent procedure with  the appropriate definition of
$d^{-1}$,
one  obtains  a 2-cocycle taking values in $\CJ$.
For example, if  $H^{3}(G)$ is trivial,
the form  ${\Tr} (g^{-1}d g)^3$ on the group $G$ is exact:
${\Tr} (g^{-1}d g)^3 = d b_2$,
where $b_2$ is a 2-form on a group $G$.
The  cocycle $C_d$ obtained  using  the descent procedure
is :
\eqn\codes{
C_d(g_1,g_2) = \lambda(g_1,g_2) + \int_{I} (\Phi(g_1)^{*} b_2+ \Phi(g_2)^{*}
b_2 -
\Phi(g_1g_2)^{*} b_2),
}
and  differs from the one presented in this paper  by a coboundary.
One can show that  the cocycle $C_d$ is independent of the
choice  of  the section  $\Phi$, while  under a change of  $b_2$ it changes by
a coboundary.

\newsec{Specializations}

Let us now assume that  $\pi_1(\CG)$  has no torsion.
We have constructed
the  universal central
extension of $\CG$ by $\CJ \oplus \pi_1(\CG) $.
All other central extensions
are formed by taking quotients
$\widehat{\CG}/\widehat{\CJ_1}$ ,where
$\widehat{\CJ_1} \subset \CJ \oplus \pi_1(\CG)$.
Here we list some interesting
examples.

First,  to get a one-dimensional central
extension we note that the group of characters of
$\CJ  \oplus  \pi_1(\CG)$ is
$Z^{n-1}_{\IZ}(X) \oplus  H^{1}(\CG)$.
Given an element $(\alpha, \chi) $ of this space we
can take $\widehat{\CJ_1}$ to be $\ker (\alpha \oplus \chi)$.
The corresponding group extension
is given by $\exp\bigl[2 \pi i \int_X \alpha\wedge  \lambda(g_1,g_2) +2 \pi i
\chi(\gamma(g_1,g_2)) \bigr]$.
Second,  if  $X$ is equipped with a metric
then we can take
$\widehat{\CJ_1} = d^* \Omega^2  +  Z^1_{\IZ}  $.
In this case  the connected component of  the center  is
the torus $H^1(X;\IR)/H^1(X;\IZ)$.
Third, if $X$ is also equipped with a complex
structure then we can work over the
complex numbers and choose
$\widehat{\CJ_1} = \bigl[ \pb^\dagger \Omega^{0,2}
\oplus  \Omega^{1,0} \bigr] +  Z^1_{\IZ} $. In this case
the connected component  of the center is  the
connected  component of  the  Picard variety
\eqn\picx{
\Pic_0(X) \equiv H^{1,0}\backslash H^1(X;\IC)/H^1(X;\IZ)
\qquad .
}
Finally, if  $X$ is a complex curve
we do not need a metric to construct
the central extension by the Jacobian
of the curve.
In this way  we obtain a solution of
the problem posed in \etingof\ and solved independently using
a very
different and
beautiful technique of holomorphic geometry in \fk.
Moreover, since  in this case $\pi_1(\CG)={\IZ}$,
the  center is the full Picard variety.

\bigskip
\bigskip
\bigskip

\centerline{\bf Acknowledgements}

We would like to thank I. Frenkel for
emphasizing to us the importance of this
problem. We would also like to thank
 L.Faddeev and  B. Khesin
for useful remarks and discussions.
The research of A. Losev was partially
supported by RFFI   grant   95-01-01101.
The research of G. Moore
is supported by DOE grant DE-FG02-92ER40704,
and by a Presidential Young Investigator Award; that of S.
Shatashvili,
by DOE grant DE-FG02-92ER40704, by NSF CAREER award and by
OJI award from DOE.

\listrefs
\bye